\documentstyle[12pt]{article}
\input tcilatex

\begin{document}

\title{Gause's exclusion principle revisited: artificial modified species and
competition }
\author{J. C. Flores$^{\dagger }$ and R. Beltran$^{\ddagger }$}
\date{$^{\dagger }$Universidad de Tarapac\'a, Departamento de F\'\i sica, Casilla
7-D, Arica, Chile.\\
$^{\ddagger }$Universidad de Tarapac\'a, Departamento de Matem\'aticas,
Casilla 7-D, Arica, Chile.}
\maketitle

\baselineskip=17pt

Gause's principle of competition between two species is studied when one of
them is sterile. We study the condition for total extinction in the niche,
namely, when the sterile population exterminates the native one by an
optimal use of resources. A mathematical Lotka-Volterra non linear model of
interaction between a native and sterile species is proposed. The condition
for total extinction is related to the initial number $M_o$ of sterile
individuals released in the niche. In fact, the existence of a critical
sterile-population value $M_c$ is conjectured from numerical analysis and an
analytical estimation is found. When spatial diffusion (migration) is
considered a critical size territory is found and, for small territory,
total extinction exist in any case. This work is motived by the
extermination agriculture problem of fruit flies in our region.

\[
{} 
\]

Published in: Jour.Phys.A:Math.Gen. 33, 4877 (2000).  


02.30 Nonlinear Differential Equations.

\newpage\ 

In ecological systems Gause's exclusion principle is widely accepted [1-5].
Originally it was deducted from competition between {\it Paramecium caudatum}
and {\it Paramecium aurelia} [1-3]. Nevertheless, it applies to many other
situations. For instance, in reference [6] was conjectured that Neanderthal
extinction in Europe was consequence of Gause's principle. In a formal point
of view, it states that {\it two competing species cannot coexist in the
same ecological niche. }In this framework, it is assumed that the strong
species fulls completely the niche and the weak disappears (exclusion). We
remark that this principle is limited in the sense that it applies when
re-adaptation, migration, or genetic changes does not exist. This principle
seems very intuitive in natural environment or for species in laboratories;
but what is the situation with genetically prepared sterile populations ?.
To be more explicit, consider the well known problem related to
extermination of native fruit flies by genetically sterilized fruit flies
[7,8]. The two species exist in the same ecological niche when the sterile
population is released. Before the interaction, we assume that the native
species fulls the niche in a stable way. In some geographic regions and for
optimal initial conditions, native fruit flies can be exterminated by the
sterile population. Namely, in this case both species disappears and the
principle must be reformulated as

\[
{} 
\]

{\it Principle of Gause :} {\it Two competing species cannot coexist in the
same ecological niche and at least one of the species disappear.}

{\it \ }

Namely, it contains explicitly the possibility of total extinction (both
species). This formulation of the principle includes all strategy of
extermination with genetically altered species [7,8]. 
\[
\]

In this paper we consider a mathematical non linear model of competition
between a native species with number of individuals $N(t),$ and other
sterile with number of individuals given by $M(t).$ Explicitly, we are
interested on the mathematical conditions for total extinction in the
ecological niche. The structure of the article is the follows: first we
present a predator-predator non linear model for the variable $N$ and $M$
(equations 1-3), and its stability analysis. Numerical simulation confirms
the stability analysis and the existence of a threshold $M_{c}$ where total
extinction exist. We give an analytical estimation of this threshold value
(equation (5)). Near to the critical value, the behavior of the extinction
time $\tau _{ext}$ is studied. This extinction time is, after our numerical
calculation, related to an critical exponent $\nu $ (equation (6)). Finally,
the case including diffusion is considered, here we found the existence of a
critical size territory $L_{c}$ were total extinction holds (for any initial
condition of the species $M$). Some possible generalizations are stressed at
the end of the paper.

\[
{} 
\]

To be explicit, consider the Lotka-Volterra type evolution equations with
interaction

\begin{equation}
\frac{dM}{dt}=-\alpha ^{\prime }M-\delta NM,
\end{equation}

\begin{equation}
\frac{dN}{dt}=NF(N)-\delta NM,
\end{equation}
where $\alpha ^{\prime }$ is the death-rate of the sterile population, $%
\delta $ is the interaction parameter, and the function $F(N)$ describes the
population growing of the native species when interaction does not exist.
For instance, when $F(N)=\alpha -N$ ($\alpha $ is a constant) we obtain the
usual Verlhust, or logistic, equation. Remark that the stability of the
point $(N=0,M=0)$ depends of the sign of $F(0).$ In fact, when $F(0)<0$ this
point becomes stable and the possibility of total extinction exists in
accordance with the Gause's principle. Moreover, this condition of stability
of $(0,0)$ seems reasonable if we think that species needs a minimal social
structure, or genetical diversity, to survival (i.e. a minimal number of
individuals). We stress that the dynamical systems (1-3) is irreversible,
for instance, a Lyapounov function $L$ associated to the systems is just $%
L=M(t)$.

\[
{} 
\]

To carry-out explicit our calculations we consider the model where

\begin{equation}
F(N)=(\alpha -N)(N-\beta ),
\end{equation}
with $\alpha $ and $\beta $ positives parameters ($\alpha <\beta $). Since $%
F(0)<0$, the point $(N=0,M=0)$ is stable and total extinction would be
expected. The linear analysis of (1-3) shows that the point $(N=\alpha ,M=0)$
is unstable (saddle) and $(N=\beta ,M=0)$ is a stable focus. Figure 1 shows
the stability diagram for our equations. So, the sterile population $M$
disappears and, depending on the initial conditions, total extinction would
exist in the niche. Namely, the systems has two atractors, the first (0,0)
related to total extinction and the other ($\beta ,0$) related to survival
of species $N$.

\[
\FRAME{itbpFU}{267.625pt}{193.75pt}{-10.0625pt}{\Qcb{Figure 1: A sketch of
the critical points associated with equations (1-2)}}{}{Figure }{\special%
{language "Scientific Word";type "GRAPHIC";maintain-aspect-ratio
TRUE;display "PICT";valid_file "T";width 267.625pt;height 193.75pt;depth
-10.0625pt;original-width 442.8125pt;original-height 302.9375pt;cropleft
"0";croptop "1.0552";cropright "1";cropbottom "0";tempfilename
'C:/SWP25/docs/FYB56O00.wmf';tempfile-properties "P";}} 
\]

The explicit question which we are concerned in this paper is the follows:
if initially the native species number is $N=\beta $ (the stable point
without interaction) then, after released $M_o$ sterile individuals, when
have we total extinction?. Namely, before the interaction, the native
species is in the niche in a stable way. After, $M_o$ sterile individuals
are released and the interaction process produces a competitive struggle.
Here we ask about the minimal population $M_o$ of sterile individuals
producing total extinction in the niche. In fact, if the sterile population
is not enough then they death and total extermination do not occur.

\[
{} 
\]

Numerical calculations confirm the existence of a critical value $M_c$ and
when $M_o>M_c$ total extinction exist in the niche. Figure 2 shows the time
behaviors of $N(t)$ for different initial value $M_o,$ of the sterile
species released in the niche. There is a critical value for the initial
condition $M_o$ related to total extinction. A criterion for total
extinction is depending on the initial number of sterile population $M_o$
and given by

\begin{equation}
M_{o}>2.7\frac{\left( \alpha -\beta \right) ^{2}}{4\delta }.
\end{equation}
This criterion is established as follows: from (1) and (3), we have that $%
M(t)=\exp \left( -\alpha ^{\prime }-\delta <N>_{t}\right) t$, where $%
<N>_{t}=(1/t)\int^{t}N(\tau )d\tau $ . When $t\rightarrow \infty ,$ and
assuming total extinction, we expect an exponential decaying behavior for $M$%
. So, an important fraction of the decaying process, assumed slow, is
reached when $M\sim M_{o}e^{-1}.$ If now we stressed that not other
stationary point (excepting $(0,0)$ exist in (2), we obtain the criterion
(4). We have used the maximum value of the function $F(N)$ given by (3). We
remark that this is a coarse criterion, nevertheless, it works in accord
with our numerical simulations. For instance, the figure 2 describes
extinction when $M_{0}>270$ in accordance with (4). This is true also for
other parameter values. 
\[
\FRAME{itbpFU}{230.0625pt}{286.0625pt}{-10.0625pt}{\Qcb{Figure 2: A
numerical simulations of equations (1-3)}}{}{Figure }{\special{language
"Scientific Word";type "GRAPHIC";maintain-aspect-ratio TRUE;display
"PICT";valid_file "T";width 230.0625pt;height 286.0625pt;depth
-10.0625pt;original-width 163pt;original-height 162.9375pt;cropleft
"0";croptop "1.0366";cropright "1";cropbottom "0";tempfilename
'C:/SWP25/docs/FYB5L302.wmf';tempfile-properties "P";}} 
\]

The criteria (4) can be generalized easily to a system with arbitrary
distribution $F(N)$ in (1-2). Namely, by imposing the inequality $F_{\max
}<\delta M_{o}e^{-1}$ (with $e\sim 2,7).$

\[
{} 
\]

The figure 3 shows a simulation of the final native population $N(t=\infty )$
for different initial condition $M_o$ of the sterile population. Clearly
there is a critical value $M_c$ which separates the survival and
extermination regime. From equation (4), a first approximation for this
critical values is

\begin{equation}
M_{c}\sim 2.7\frac{(\alpha -\beta )^{2}}{4\delta }.
\end{equation}

\[
\FRAME{itbpFU}{233.3125pt}{243.25pt}{-10.0625pt}{\Qcb{Figure 3: The final
distribution of native population $M$}}{}{Figure }{\special{language
"Scientific Word";type "GRAPHIC";maintain-aspect-ratio TRUE;display
"PICT";valid_file "T";width 233.3125pt;height 243.25pt;depth
-10.0625pt;original-width 163pt;original-height 162.9375pt;cropleft
"0";croptop "1.0433";cropright "1";cropbottom "0";tempfilename
'C:/SWP25/docs/FYB5OK03.wmf';tempfile-properties "P";}} 
\]

Moreover, figure 2 suggests that near to this critical value the extinction
time $\tau _{ext}$ depends on $(M_o-M_c)$. This is so because when $%
M_o\rightarrow M_c^{+}$ the extinction time must be infinity. Explicitly we
expect a behavior like

\begin{equation}
\tau _{ext}\sim \frac 1{\left( M_o-M_c\right) ^\nu };\text{ }M_o>M_c,
\end{equation}
where $\nu $ is a unknown parameter. The evaluation of this critical
exponent requires a computational work beyond of the scope of this paper. It
will be do elsewhere. The conjecture (6) is reinforced by numerical
calculation. In fact, using the same parameter values that in figure 2, and
the definition

\begin{equation}
\tau _{ext}^{-1}=-\frac 1t\lim _{t\rightarrow \infty }Ln(N(t)/\beta ),
\end{equation}
we see that the existence of the critical exponent $\nu $ is confirmed
numerically (figure 4).

\[
\FRAME{itbpFU}{230.75pt}{240pt}{-10.0625pt}{\Qcb{Figure 4: The extinction
time $\protect\tau _{ext}$}}{}{Figure }{\special{language "Scientific
Word";type "GRAPHIC";display "PICT";valid_file "T";width 230.75pt;height
240pt;depth -10.0625pt;original-width 163pt;original-height
162.9375pt;cropleft "0";croptop "1.0440";cropright "1";cropbottom
"0";tempfilename 'C:/SWP25/docs/FYB52204.wmf';tempfile-properties "P";}} 
\]

Now we discuss briefly the incorporation of migration to the set of
evolution equations (1-2). In fact, total extermination could be also
carried out by diffusion process. In some cases unstable points become
stable by diffusion in limited territories. Thus in a model where (0,0) is
unstable, i.e. only one species survive, diffusion would change this
instability and total extinction takes place. We add the diffusion terms $D_M%
\frac{\partial ^2M}{\partial x^2}$ and $D_N\frac{\partial ^2N}{\partial x^2}$
to (1) and (2) respectively. Namely, consider the pair of reaction-diffusion
evolution equations

\begin{equation}
\frac{dM}{dt}=-\alpha ^{\prime }M-\delta NM+D_M\frac{\partial ^2M}{\partial
x^2},
\end{equation}
\begin{equation}
\frac{dN}{dt}=NF(N)-\delta NM+D_N\frac{\partial ^2N}{\partial x^2}.
\end{equation}
The linear analysis of stability for the stationary point $(0,0)$ can be do
in the usual way, namely, consider the small perturbation 
\begin{equation}
M=0+\eta ,
\end{equation}

\begin{equation}
N=0+\varepsilon ,
\end{equation}
where the variables $\eta $ and $\varepsilon $ are assumed small. Equations
(8) and (9), give the first order linear equations

\begin{equation}
\frac{\partial \eta }{\partial t}=-\alpha ^{\prime }M\eta +D_M\frac{\partial
^2\eta }{\partial ^2x},
\end{equation}
\begin{equation}
\frac{\partial \varepsilon }{\partial t}=F(0)\varepsilon +D_N\frac{\partial
^2\varepsilon }{\partial ^2x},
\end{equation}
where we assume $F(0)>0,$ corresponding to the unstable case when migration
is not present. For finite territory, solutions like $\varepsilon \sim
e^{\omega t}\sin kx$ can be visualized. The relationship between the
stability parameter $\omega $ and the wavenumber $k$ is given by

\begin{equation}
\omega =F(0)-k^2D_N,
\end{equation}
and clearly for $k>\sqrt{F(0)/D_N}$ the point $(0,0)$ becomes stable and
total extinction in the niche exist. Since $k\sim \frac 1L$, with $L$ the
territory size, equation (14) defines a critical size territory $L_c\sim 
\sqrt{D_N/F(0)}$ where total extinction holds. Namely, for all size
territory $L<L_c$ total extinction exist.

\[
{} 
\]

In conclusion: the principle of Gause was generalized to consider the case
of biological struggle when one competing species is sterile. In fact, under
appropriate conditions, total extinction could occur in the niche. Most of
the agriculture competitive extermination method are carried-out assuming
this principle. For instance, this is the case of extermination program of
fruit flies with sterile flies irradiated in laboratories [9-10]. We have
presented a simple model which has total extinction in the niche in some
cases. Conjectures related to a critical sterile-population (number of
individuals) producing total extinction were pointed-out with a coarse
criterion (4). This conjectures are based in our numerical simulation of the
model. The role of migration was briefly discussed and the possibility of
total extinction from diffusion was also explored for small territories.

\[
\]

To ending, we note that our model can be extended to incorporates some
modifications. Particularly we are thinking about generalizations like:

(i) Periodic variation of coefficients. In fact, in the extinction fruit
flies programs, daily, season or El ni\~{n}o (ENSO), oscillations must be
considered.

(ii) Sexual selection. Many extermination programs are based on sexual
selection, namely, sterile-male released in a given niche. It creates
interaction between sterile-male and fertile-female which becomes directly
related to the evolution of the native-male. Such models require a
phase-space greater than two.

(iii) Many random factors are present in real niche. For instance, humidity,
temperature, wind, etc. These factors can be incorporated in our model by
introducing adequate sthocastic process for the parameter ($\alpha ,$ $\beta
,$ or $\delta $).

\[
{} 
\]

Acknowledgment: we thanks Hernan Donoso (S. A. G. Arica) for introduce us at
the problem of extinction of fruit flies with sterile populations. We thanks
C. Romo for the revision of the manuscript. J. C. F. thanks E. Martin and C.
Saravia for the initial help concerning the subject.

\

\end{document}